\documentclass{article}
\usepackage[latin9]{inputenc}
\usepackage{color}
\usepackage{amsmath}
\usepackage{graphicx}
\usepackage{subcaption}
\makeatletter
\newcommand{\ba}{\begin{array}}
\newcommand{\ea}{\end{array}}
\newcommand{\bc}{\begin{center}}
\newcommand{\ec}{\end{center}}
\newcommand{\be}{\begin{equation}}
\newcommand{\ee}{\end{equation}}
\newcommand{\hsp}{\hspace{1cm}}
\newcommand{\dd}{\varphi}
\newcommand{\w}{\psi}
\newcommand{\la}{\lambda}
\newcommand{\al}{\alpha}
\newcommand{\bb}{\beta}

\newcommand{\veta}{\vec{\eta}}

\newcommand{\vxi}{\vec{\xi}}
\newcommand{\vr}{\vec{r}}

\newcommand{\rij}{r_{ij}}

\newcommand{\esp}{\vspace{.1cm}}
\newcommand{\beq}{\begin{equation}}
\newcommand{\eeq}{\end{equation}}
\newcommand{\beqa}{\begin{eqnarray}}
\newcommand{\eeqa}{\end{eqnarray}}
\newcommand{\bdis}{\begin{displaymath}}
\newcommand{\edis}{\end{displaymath}}
\begin{document}
\vspace{2.in}
\begin{center}
\begin{Large}
\begin{bf}
Analytical Expressions for a Hyperspherical Adiabatic Basis \\
Suitable for a particular Three Particle Problem \\
in 2 Dimensions \\
\end{bf}
\end{Large}
\vspace{.7cm}
 Monique Lassaut \\
Universit\'e Paris-Saclay,CNRS/IN2P3, IJCLab, 91405 Orsay, France\\
 Alejandro Amaya-Tapia\\
 Instituto de Ciencias F\'{i }sicas, Universidad Nacional Aut\'onoma de M\'exico,\\
 Av. Universidad s/n, Col. Chamilpa, Cuernavaca, Mor. 62210, M\'{e}xico.\\
Anthony  D. Klemm \\
School of Computing \& Mathematics, Deakin University, Geelong, Victoria,
Australia \\
{\it and} \\
Sigurd Yves Larsen \\
Department of Physics, Temple University, Philadelphia, PA 19122, USA  \\

\vspace{1.cm}
\begin{bf}
ABSTRACT \\
\end{bf}
\vspace{.3cm}
\end{center}
For a particular case of three-body scattering in two dimensions, 
and matching analytical expressions at a transition point, we obtain 
accurate solutions for the hyperspherical adiabatic basis and potential.
We find analytical expressions for the respective, asymptotic, inverse 
logarithmic and inverse power potential behaviours, that arise as functions 
of the radial coordinate.  
The model that we consider is that of two particles interacting with a 
repulsive step potential, a third particle acting as a spectator. The model 
is simple but gives insight, as the 2-body interaction is long ranged in 
hyperspherical coordinates. The fully interacting 3-body problem is known, 
numerically, to yield similar behaviours that we can now begin to understand.
That, clearly, is the ultimate aim.
\newpage
\topmargin=0in
\pagestyle{plain}
\pagenumbering{arabic}
\setcounter{page}{2}
\begin{center}
\section*{Introduction}
\end{center}
\vspace{.3cm}

In a previous paper \cite{klemm}, the authors showed how, starting from
hyperspherical harmonic expansions, they obtained
adiabatic potentials, suitable for the calculations of three-body
phase shifts at low energies. They constructed matrices consisting of 
hyperspherical harmonic matrix elements of the potential, together with
centrifugal terms. The matrices were then diagonalized, to yield
the desired adiabatic potentials. 
The calculations were for 3 particles
in a plane, subject to finite repulsive core interactions.

\esp
The calculations were meant to establish a method which would lead to
the evaluation, at low temperature, of a third fugacity coefficient
in Statistical Mechanics.
The latter task was subsequently carried out by Jei Zhen
and one of the authors \cite{zhen}.
In both investigations, it was important to consider different
cases, corresponding to the distinct representations of the permutation group
and different physical situations, with either the 3 particles interacting or
simply two of them interacting, with the third acting as a spectator.

\esp
Absolutely crucial, in these investigations, is the large-$\rho$ behaviour
of the effective potentials (the adiabatic eigenvalue
minus the appropriate centrifugal term). 
The nature of the long ``tail'' of the effective
potential determines how the correspondent eigenphase shift behaves,
as the energy tends to zero. Thus, our most significant result was that for
the 3 most important types of the phase shifts,
associated with the cases of ${}^{0}\Gamma_{1g}$, ${}^{0}\Gamma_{2g}$
and $\overline{\delta}$, the effective potentials 
behave as $1/(\rho^2 \ln \rho)$, for large values of
$\rho$, instead of the $1/\rho^2$ of the hyperspherical potential matrix
elements. [The potential matrix elements are polynomials in $1/\rho^2$.]
This then implies that the phase shifts, instead of tending to constants
as the energy goes to zero, behave as $1/(\ln\,q)$, and therefore
go to zero! (The variables $\rho$ and $q$ are, respectively, the hyper
radius and the reduced wave number.) Other phase shifts were found to go to 
zero more powerfully. 
We show in Fig. 1 an example of the behaviour of the effective potential 
in the case ${}^{0}\Gamma_{1g}$ .


\vspace{1cm}
\begin{figure}[h]
\centering{}\centering{}
\includegraphics[scale=0.42] {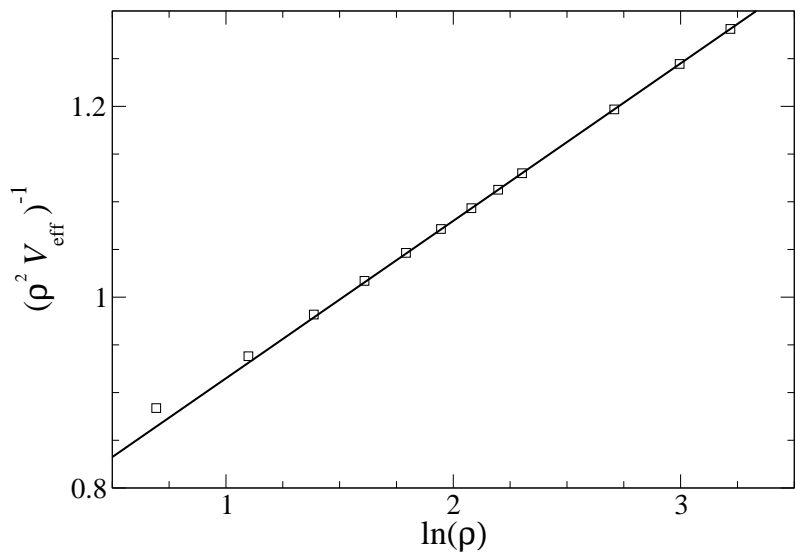}
\caption
{This figure shows the logarithmic behaviour for $\rho$ large of the effective 
potential, as a function of $\rho$, $\rho$ being the hyperradius, 
in the symmetry ${}^{0}\Gamma_{1g}$. Both, the calculated data (square 
symbols) and a fitting to the data (straight line), were taken from 
\cite{zhen}
}
\end{figure}

\newpage

\esp
Though, in our old paper, our basic material was numerical,
we were able nevertheless to propose ``heuristic'' formulae, to characterize
the asymptotic behaviour of the $3$ types of eigenpotentials, of the remodelling
that takes place to yield a different scattering from the one expected from
the solution of a finite number of hyperspherical equations.

\esp
In this paper, we show that in one of
the three cases mentioned above, the case $\overline{\delta}$, we succeed
in calculating accurately the adiabatic eigenvectors and eigenvalues for all
the values of $\rho$, and present analytical expressions valid  in the 
asymptotic 
region.

\esp
While this calculation involves a case where only two of the particles
interact, while the 3rd particle acts as a spectator, it is well to note that
in the hyperspherical coordinate system the two-body interaction is
long ranged (in $\rho$) and also that in the full hyperspherical
calculations
of the other cases, we only need, using symmetry and enforcing a restriction
on the quantum numbers, to
take into account the matrix element of one of the pair potentials.

\esp
Here, then, our calculations allow us to re-examine our previous results, and
confirm and extend the asymptotic forms (and coefficients) that can be used
to characterize the long range behaviour of the various effective  potentials.

\newpage
\begin{center}
\section*{The KL Hyperspherical Coordinate System}
\end{center}
\subsection*{The Harmonic Basis}
\vspace{.2cm}

For a system of three equal mass particles in two dimensions, 
 we define the Jacobi coordinates
\[
 \veta = (\vr_1 - \vr_2)/\sqrt{2} \hsp \mbox{and} \hsp \vxi = 
 \sqrt{2/3}\, \left(\frac{\vr_1+\vr_2}{2} - \vr_3 \right)\;,
\]
which allows us to separate, in the Hamiltonian, the center of mass
coordinates from those associated with the internal motion.

\esp
Kilpatrick and Larsen \cite{john} then introduce hyperspherical coordinates,
associated with the moment of inertia ellipsoid, of the $3$ particles,
which allows them to disentangle permutations from rotations and obtain
harmonics which are pure representations of both the permutation
and the rotation group. Taking the $z$ axis normal to the plane of
the masses, we write for the cartesian components of the Jacobi coordinates
\beqa
\eta_{x} & = & \rho (\cos \vartheta \cos \varphi \cos  \psi
                    + \sin \vartheta \sin \varphi \sin \psi ),  \nonumber \\
\eta_{y} & = & \rho (\cos \vartheta \cos \varphi \sin  \psi
                    - \sin \vartheta \sin \varphi \cos \psi ),  \nonumber \\
 \xi_{x} & = & \rho (\cos \vartheta \sin \varphi \cos  \psi
                    - \sin \vartheta \cos \varphi \sin \psi ),  \nonumber \\
 \xi_{y} & = & \rho (\cos \vartheta \sin \varphi \sin  \psi
                    + \sin \vartheta \cos \varphi \cos \psi ),
\eeqa
in terms of
the hyper radius $\rho$ and of the three angles $\vartheta$, $\varphi$ and
$\psi$.

\esp
The harmonics, in their unsymmetrized form, are then 
\be
Y_N^{\nu \la}(\Omega) = N_n^{\al \bb} \Theta_n^{\al \bb}(x) e^{i\nu\dd}
                        e^{i\la\w} 
\label{2} \;,
\ee
   where $x = \sin{2\vartheta}$ and
\be
  \Theta_n^{\al \bb}(x) = (1 - x)^{\al/2}(1 + x)^{\bb/2} P_n^{\al,\bb}(x) \ .
\ee
$P_n^{\al,\bb}(x)\;$ is a Jacobi polynomial, and the
 normalization constant is
\[
 N_n^{\al \bb} = \left\{\left(\frac{N + 1}{2^{\al +\bb + 1}}\right)
              \left( \ba{c} n + \al + \bb \\  \al \ea \right)
              \left( \ba{c} n + \al \\ \al \ea \right)^{-1} \right\}^{1/2} \;.
\]
   The hyper radius $\rho\;$ satisfies $\rho^2 = \eta^2 + \xi^2$
and the angular components have the ranges
\[
 -1 \leq x \leq 1, \hsp -\pi/2 \leq \dd \leq \pi/2,
 \hsp 0 \leq \w \leq 2\pi\;.
\]
   Finally we have for the indices the relations
\[
 n = \frac{1}{2}[N - \mbox{{\rm max}}\{| \nu|,|\la|\}], 
 \hsp \al = \frac{1}{2}|\nu +\la|, \hsp \bb = \frac{1}{2}|\nu - \la|\;,
\]
   where $N$ is the degree of the harmonic, and $\la$ is the inplane
   angular momentum quantum number. The indices 
 $\nu$ and $\la$ take on the values $-N$ to $N$ in steps
of $2$; all three have the same parity and $N = 0,1,2,\ldots.$

\esp
Linear combinations of the basic harmonics can then be formed \cite{john} 
to obtain irreducible bases, adapted to the symmetries of the
physical problems \cite{klemm,zhen}.
\newpage
\subsection*{The Adiabatic Basis}
\vspace{.2cm}

For our model, the particles interact via a binary step potential
\be
     V(\rij)  =  \left\{ \ba{ll}
                          V_0, & \rij \leq \sigma \\
                          0, & \rij > \sigma
                       \ea
                 \right. \hsp  
\ee
where the height $V_0$ and the range $\sigma$, are both finite.

\esp
In order for the $\lambda$ in the following equation, to be the 
same as that used by \cite{klemm} and \cite{zhen}, 
we need to add a term $3/(4\rho^2)$.
That is because the angular part of the Laplacian has eigenvalues
$- N(N+2)$ and not the $ -[(N+1)^2 - 1/4]$ used in the coupled equations
used by these authors.

\esp
The adiabatic eigenfunctions $B_{\ell}$ are then defined as satisfying
\be
 \left\{ - \frac{1}{\rho^2}\nabla_{\Omega}^2 + \frac{3}{4 \rho^2} +
     \frac{2m}{\hbar^2} V(\rho,\Omega) \right\} B_{\ell}
     = \lambda_{\ell}(\rho) B_{\ell} \;, 
\ee
where $ V(\rho,\Omega)$ is either the sum of the binary potentials or,
simply one of the binary potentials, say $V(r_{12})$, expressed as a function 
of $\rho$ and the angles. The index $\ell$ stands for the set of quantum
numbers which characterize and index the particular class of solutions.
$\lambda_{\ell}(\rho)$ is the eigenvalue, which upon subtraction of a 
``centrifugal'' type term yields the effective potential, of concern to
us later on.

\esp
The eigenfunctions may now be used to expand the wavefunctions of
the physical systems:
\be
\Psi = \sum_{\ell^{\prime}} B_{\ell^{\prime}}(\rho,\Omega)
\phi_{\ell^{\prime}}(\rho) \;, 
\ee
where the amplitudes $\phi_{\ell}(\rho)$ are the solutions of the
coupled equations:
\beq
-\sum_{\ell^{\prime}}\int\! d\Omega \,B^{\ast}_{\ell}
(\rho,\Omega)\frac{\partial
^{2}}{\partial\rho^{2}}\left(B_{\ell^{\prime}}(\rho,\Omega)\phi_{\ell^{
\prime}}(\rho)\right) + \lambda_{\ell}(\rho)\phi_{\ell}(\rho)
= (2mE/\hbar^{2})\phi_{\ell}(\rho) .
\eeq

The adiabatic eigenfunctions can themselves be expanded in hyperspherical
harmonics and this is how a large set of them were calculated in the 
papers quoted earlier.
The symmetries of the hyperspherical harmonic basis are, of course,
reflected in the solutions of the adiabatic eigenvectors. For the fully
symmetric Hamiltonian, the set of solutions divides into six separate
subsets \cite{john}, each requiring calculations involving combinations of 
matrix elements of only one of the binary potentials, but with restrictions
on the quantum numbers of the unsymmetrized harmonics involved.
In the case of two interacting particles, with a third as a spectator,
we find an additional four subsets. 

\esp
The numerical approach was then, for each $\rho$, to evaluate a large 
potential matrix, with the appropriate harmonic basis, add to this the
(diagonal) ``centrifugal'' contribution arising from the angular part of
the kinetic energy (the angular part of the Laplacian in the Hamiltonian)
and diagonalize to obtain the required adiabatic eigenvalues. The
number of harmonics, needed for numerical convergence, increases as a function
of $\rho$, but it was our fortunate experience to find that it was possible
to evaluate correctly the eigenvalues, that we sought, for values of
$\rho$ large enough that the behaviour of $\lambda_{\ell}(\rho)$ could
be described by asymptotic forms. We were able to characterize them, and 
this gave us the values of $\lambda_{\ell}(\rho)$ for all the larger
values of $\rho$.

\newpage
\begin{center}
\section*{Dual Polar Set of Coordinates}
\end{center}
\subsection*{The Harmonic Basis}
In this part of the paper we wish, exclusively, to consider the case of
two particles interacting together, the third acting as a spectator.
As we shall show, we are then able to obtain exact adiabatic solutions.

Our reasoning is as follows.
When the third particle does not interact with the other two, this must 
imply that the motion of the pair (1,2), and therefore its angular momentum,
is unaffected by the motion of the third particle. In a parallel fashion,
the motion of the third particle, and its angular momentum about the center
of mass of the particles (1,2), must be a constant as well.
If we choose our coordinates carefully, the angular behaviour of two of the
angles should ``factor'' out and, for a given $\rho$, only one variable should
be involved in a key differential equation.

We note that in the KL coordinates, the distances between particles involve
two of the angles, for example $r_{12}^2$ equals $\rho^2 (1 + \cos 2\vartheta
\cos 2\varphi)$.  To get around this, we choose an angle to give us the 
ratio of the length of the $2$ Jacobi vectors, and then polar coordinates
for each of them.
Thus, we represent $\Omega$ by $(\theta_1, \theta_2, \phi)$, where 
 $\eta = \rho \cos \phi$, $\xi = \rho \sin \phi$ and $\eta_x = \eta \cos
\theta_1$,  $\eta_y = \eta \sin \theta_1$,  $\xi_x = \xi \cos \theta_2$,
 $\xi_y = \xi \sin \theta_2$. The ranges of these angles are
\[
   0 \leq \phi \leq \pi/2, \hsp 0 \leq \theta_1 \leq 2\pi, \hsp 0 \leq
\theta_2 \leq 2 \pi \;.
\]

To obtain the harmonics, in a manner which is suitable to also demonstrate the 
link with the KL harmonics, we introduce complex combinations of the Jacobi
coordinates, i.e. the monomials
\beqa
z_{1} & = &  (\eta_{x} + \imath \eta_{y}) 
                     \nonumber \\
z^{*}_{1} & = &  (\eta_{x} - \imath \eta_{y})  \nonumber \\
z_{2} & = &  (\xi_{x} + \imath \xi_{y})
                     \nonumber \\
z^{*}_2 & = & (\xi_{x} - \imath \xi_{y})
\eeqa
It then follows that 
\beqa
\rho^2 & = & (z_{1} z^{*}_{1} + z_{2} z^{*}_{2}) \nonumber \\
\nabla^{2} & = & 4 \, \left(\frac{\partial^{2}}{\partial z_{1} \partial z^{*}_{1}}
   + \frac{\partial^{2}}{\partial z_{2} \partial z^{*}_{2}}\right),
\eeqa
and, clearly, $z_{1}$, $z^{*}_{1}$, $z_{2}$ and $z^{*}_2$ each satisfies
Laplace's equation, as do the combinations $z_{1}z_{2}$, $z_{1}z^{*}_{2}$,
$z^{*}_{1}z_{2}$, $z^{*}_{1}z^{*}_{2}$ and these combinations raised to 
integer powers. 

Writing $\rho_{1}^{2} = z_{1}z^{*}_{1}$
and $\rho_{2}^{2} = z_{2}z^{*}_{2}$, we can write as the most general solution
arising from the monomials $z_{1}$ and $z_{2}$:
\bdis
z_{1}^{\ell_{1}} z_{2}^{\ell_{2}} P_{\ell}^{\ell_{2},\ell_{1}}\left(\frac{\rho_{2}^{2} -
\rho_{1}^{2}} {\rho_{2}^{2} + \rho_{1}^{2}}\right) ( \rho_{1}^{2} +
\rho_{2}^{2} )^{\ell}\, ,
\edis
where $\ell_1$, $\ell_2$ and $\ell$ are positive integers or zero, and 
$P_{\ell}^{\ell_{2},\ell_{1}}$ is a Jacobi polynomial. \\ 
\noindent
In terms of the angles, our expression becomes proportional to:
\bdis
\rho^{\ell_{1} + \ell_2 + 2\ell } (\cos^{2} \phi)^{\ell_{1}/2} (\sin^{2} \phi)^{\ell_{2}/2}
P_{\ell}^{\ell_{2},\ell_{1}}(\cos 2 \phi) e^{\imath\theta_{1}\ell_{1}} 
e^{\imath\theta_{2}\ell_{2}} \, ,
\edis
and, finally, in terms of $z$ equal to $\cos 2\phi$, we define our 
unnormalized harmonic:
\beq
Y_{\ell}^{\ell_{1},\ell_{2}}(\theta_{1}, \theta_{2}, z) = 
(1 + z)^{|\ell_{1}|/2}(1 - z)^{|\ell_{2}|/2}\,
P_{\ell}^{|\ell_{2}|,|\ell_{1}|}
(z)\, e^{\imath\theta_{1}\ell_{1}} e^{\imath\theta_{2}\ell_{2}}\, ,\label{ya}
\eeq
where now $\ell_{1}$ and $\ell_{2}$ can be positive, negative, integers - or zero.
(This takes into account the other combinations $z_{1}z_{2}^{*}$, etc.)
The order of the harmonic is $N$ equal to $|\ell_{1}| + |\ell_{2}| + 2\ell$. 
\vspace{.5cm}
\subsection*{The Adiabatic Differential Equation}

Writing
\beq
\nabla_{\eta}^{2} + \nabla_{\xi}^{2} = \left(\,\frac{\partial^{2}}
{\partial\rho^{2}} + \frac{3}{\rho}\frac{\partial}
{\partial\rho}\,\right) + \frac{1}{\rho^2}\nabla_{\Omega}^2\, ,
\eeq
inserting our polar coordinates into the left hand side and changing to our
variable $z$, we find:
\beq
 \nabla_{\Omega}^2 = 
4 (1 - z^2) \frac{\partial^2}{\partial z^2} - 8 z \frac{\partial}{\partial z}
+ \frac{2}{(1+z)}\frac{\partial^2}{\partial\theta_{1}^{2}}
	      + \frac{2}{(1-z)}\frac{\partial^2}{\partial\theta_{2}^{2}}\,\,.
\eeq
If we now write our adiabatic eigenfunctions as
\beq
 B_{N}^{\ell_1 \ell_2}(\rho, \Omega) = e^{i \ell_{1} \theta_1} e^{i \ell_2 \theta_2}\,
 (1 + z)^{|\ell_{1}|/2}(1 - z)^{|\ell_2|/2}\,  F_{\ell}^{|\ell_1|,|\ell_2|}(\rho,z)\;, \label{yb}
\eeq
then the functions $F$ will satisfy the equation:
\beqa
& &  \left[ - 4 (1 - z^2) \frac{\partial^{2}}{\partial z^{2}}  + 
4((2 + \ell_1 + \ell_2)z + \ell_2 - \ell_1)\frac{\partial}{\partial z} \right] 
  F_{\ell}^{\ell_1,\ell_2}(\rho, z)  \nonumber \\
& + &   \left[ (\ell_1 + \ell_2)(\ell_1 + \ell_2 + 2) + \frac{3}{4} +
  \rho^2 \overline{V}(\rho,z) \right] \ \ F_{\ell}^{\ell_1,\ell_2}(\rho, z)
 =  
\rho^2 \lambda(\rho) F_{\ell}^{\ell_1,\ell_2}(\rho, z) 
\label{14}
\eeqa
where $\overline{V}(\rho,z)$ equals $2m/\hbar^{2}$ times the potential and
in our notation we have dropped the absolute value indications. All the indices will be understood to be positive or zero. 

\esp
When $\overline{V}(\rho,z) = 0$, we can  obtain a 
solution which is analytic between $-1 \leq z \leq +1$. 
For $\lambda$ equal to $ (\ell_1 + \ell_2 + 2 \ell)(\ell_1 + \ell_2 + 2\ell + 2)/\rho^2$ and
$\ell$ a non-negative integer, we 
find that our $F$ is simply $P_{\ell}^{\ell_{2},\ell_{1}}(z)$, the Jacobi polynomial
which appears in our Eq. (\ref{ya}).
The $N$ that appears in the $B$ of Eq. (\ref{yb}) is the order of the
corresponding harmonic. \\
We now scale our $\rho$. I.e., we let our new $\rho$ equal our old 
$\rho/\sigma$. Then, for $\rho > \frac{1}{\sqrt{2}}$ and our new potential
\be
    \overline{V}(\rho,z) = \left\{ \ba{ll}
   (2m/\hbar^2) V_{0} \,\sigma^2 & -1 \leq z \leq -1 + 1/\rho^2 \\
               0  & -1 + 1/\rho^2 < z \leq 1 \;,
                     \ea \right. 
\ee
the solutions of this equation, which behave reasonably at 
$z$ equal to $-1$ and $+ 1$, will be seen to be
proportional to extensions of the Jacobi polynomials to functions with
non-integer indices, in a relationship similar to that of Legendre 
polynomials and Legendre functions.
To motivate and clarify our
procedure we first consider the case of $\ell_1 = \ell_2 = 0$, with and without
potential.

When the potential is put to zero and we factor a $4$ as well as change the 
sign, the differential equation reads
\beq
  \left[(1 - z^2) \frac{\partial^{2}}{\partial z^{2}}  - 2z 
 \frac{\partial}{\partial z} + \ell\,(\ell + 1)\right] \; F_{\ell}^{0,0}(\rho, z) = 0.
\eeq
This is, of course, the Legendre differential equation and, with $\ell$ a positive
or zero integer, the well behaved solutions are the Legendre polynomials.

In the case of our potential, which is zero or a constant (only a function of
$\rho$) in the different ranges of $z$, we can write our differential equation
in a very similar form, i.e. as
\beq
\left[(1 - z^2) \frac{\partial^{2}}{\partial z^{2}}  - 2z
\frac{\partial}{\partial z} + \nu\,(\nu + 1)\right] \; F_{\nu}^{0,0}
(\rho, z) = 0 \,,
\eeq
where for $-1+1/\rho^2 < z \leq 1$
\beq
\hspace{-1.4cm}\nu\,(\nu + 1) = \rho^2\,\lambda\,(\rho)/4 - \frac{3}{16}
\eeq
and\, for $-1 \leq z \leq -1+1/\rho^2$
\beq
\nu\,(\nu + 1) = \rho^2\,[ \lambda\,(\rho) - \overline{V}_0]/4 - 
\frac{3}{16}\;. 
\eeq

Denoting the respective values of $\nu$ as $\nu_1$ and $\nu_2$, the
corresponding solutions are the Legendre function $P_{\nu_1}(z)$ and the
combination 
\bdis
 P_{\nu_{2}}(-z) = \cos(\pi \nu_{2})\, P_{\nu_{2}}(z) - 
(2/\pi)\, \sin(\pi \nu_{2})\, Q_{\nu_{2}}(z)\,,
\edis
of the first and second Legendre functions.

\esp
The point is as follows. Whereas $P_{\nu_{1}}(z)$ is well behaved at
$z$ equal to $1$, and is suitable as a solution for its
range in $z$ from $-1 + 1/\rho^2$ to $1$, both the 
$P_{\nu_{2}}(z)$ and $Q_{\nu_{2}}(z)$ have a logarithmic singularity at $z$ 
equals $-1$. The combination that we propose, however, is such that the
logarithmic terms cancel out and the combination \cite{abram} is a well
behaved solution in the range  $-1$ to $-1 + 1/\rho^2$.

Expressing these solutions as power series, the first about $z$ = $1$,
the second
about $z$ = $-1$, we obtain 
\begin{eqnarray}
 {}_2F_1(-\nu_1, \nu_1 + 1; 1;
    \frac{1}{2}(1-z)), & \mbox{for}   -1+1/\rho^2 < z \leq 1 \nonumber\\
   \mbox{and} &     \hsp \label{twin} \\
 {}_2F_1(-\nu_2, \nu_2 + 1; 1;
       \frac{1}{2}(1+z)), & \mbox{for}  -1 \leq z \leq -1+1/\rho^2 \nonumber
\end{eqnarray}
Our overall solutions are then obtained by matching the logarithmic
derivative of the two solutions (above) at the boundary: 
at $z$ equal $-1 + 1/\rho^2$. 
This then also yields the adiabatic eigenvalues.

It now remains to note that for the cases of $\ell_1$ and $\ell_2$ not 
equal to zero, we can use the same procedure. We have, for the two regimes,
solutions proportional to 

\begin{eqnarray}
{\rm (R)} & _{2}F_{1}(-\nu_{1},\nu_{1}+|\ell_{1}|+|\ell_2|+1;|\ell_2|+1;\frac{1}{2}(1-z)), & \mbox{for}-1+1/\rho^{2}<z\leq1 \label{eq11}\\
 & \hspace{1cm} & \mbox{and} \nonumber\\
(\mathrm{L}) & _{2}F_{1}(-\nu_{2},\nu_{2}+|\ell_{1}|+|\ell_2|+1;|\ell_{1}|+1;\frac{1}{2}(1+z)), & \mbox{for}-1\leq z\leq-1+1/\rho^{2} \label{eq21}
\end{eqnarray}
henceforth labelled Right $\left(R\right)$ and Left $\left(L\right)$.

 For each choice of $\ell_1$ and $\ell_2$
there is an infinite set of values of $\nu_1\;$ for which the logarithmic
derivative of the hypergeometrical  functions can be matched at $z$ equal to 
$-1 + 1/\rho^2$. For each such value of $\nu_1$, the 
adiabatic eigenvalue is then given by
\be
  \lambda (\rho) = \frac{(2 \nu_1 + |\ell_1| + |\ell_2| + 1)^2 - \frac{1}{4}}
                              {\rho^2} \;. 
\ee


When $V_0 = 0$, the adiabatic basis
reduces to the hyperspherical harmonic basis of Eq. (\ref{ya}), since
 the hypergeometrical  functions reduce to Jacobi polynomials, and
 $\nu_1 \equiv \nu_2 = \ell$. So our $B_{N}^{\ell_1,\ell_2}$ is precisely the 
$Y_{\ell}^{\ell_1,\ell_2}(\theta_1,\theta_2,z)$.
\newpage
\section*{Comparison of the Adiabatic Eigenvalues}
First a historical note. 

\esp
\noindent
The results obtained by diagonalizing matrices to obtain the adiabatic 
potentials, and shown here below in this section, were drawn from 
reference \cite{jei}. The `direct' results 
are obtained by solving numerically Eq. (\ref{14}), matching the
logarithmic derivatives of the appropriate analytical solutions, at the
edge of the binary potential.

\esp
When the numerical work was done (using the KL basis),
lists were made of the appropriate harmonics needed to form the
matrices (potential and centrifugal) which, when added and diagonalized,
yield the adiabatic eigenvalues. We now need to identify these
eigenvalues and compare them with those obtained by the new method. This
is not trivial, but an immediate remark can be made.

\esp
First of all, in both approaches the angular momentum $\la$ is a good 
quantum number. Further, in the new basis, we can write:  
\be
 \la = \ell_1+ \ell_2 \;. 
\ee
 This follows from the fact that $\ell_1$ specifies the angular momentum of
the 1-2 pair and $\ell_2$ specifies the angular momentum of the third particle
relative to the center of mass of the first two. Thus their sum defines the
total inplane angular momentum. Hence, for example, when $\la = 0$ we can
have all pairs $\ell_1$ and $\ell_2$ with $\ell_1 = - \ell_2$. If $\ell_1 = \ell_2 = 0$,
this then provides a single eigenvalue for each choice of $N = 2\,\ell,\,
\ell = 0,1,2,\ldots.$

\esp
Another indicator is whether $n$ is even or odd, which is very significant in
the  drawing up of the lists, associated with the symmetries of the harmonics.
Proceeding, then, we compare values of the effective potential, defined by 
\be
  V_{{\rm eff}}(\rho, N) = \la(\rho) - \frac{(N+1)^2-\frac{1}{4}}{\rho^2}\;,
\label{22}
\ee
where we subtract from each eigenvalue the value of the centrifugal term 
that would correspond to it, if the binary potential were allowed to go to
zero.
These have been extensively tabulated by Zhen \cite{jei}, but see also 
\cite{klemm,zhen}.

\esp
  In {\em Table 1} we confirm a central result of the previous authors' work.
We demonstrate the convergence of the truncated matrix method to the result
obtained directly. This, for the simplest case, $N = 0$ and, a sample value of 
$\rho = 5\;$ and $\Lambda^* = 10$.
($\Lambda^* = (h^2/mV_0\sigma^2)^{1/2}$). As we see, the result is excellent.

\begin{table}[h]
\bc
\begin{tabular}{|cc|} \hline
$N_{{\rm max}}$ & $V_{{\rm eff}}(5, 0)$ \\ \hline
 $110$ & $0.011754744$ \\
 $120$ & $0.011754730$ \\
 $130$ & $0.011754670$ \\
 $140$ & $0.011754666$ \\ 
 Direct & $0.011754562$ \\ \hline
\end{tabular} 
\ec
\caption{\it Convergence of the matrix method}
\end{table}

\esp
  A more extensive set of comparisons is made in {\em Table 2}, where selected
values of the effective potential, obtained from eigenvalues of the truncated
matrix, are chosen for various values of $N$, $\la$ and $n$ and compared
with the direct results. In all cases, except the first,
the matrix was truncated at $N_{{\rm max}} = 100$, where $N_{{\rm max}}$ is the maximal order of the hyperspherical elements used in constructing the matrix.

 \newpage
\begin{table}
\bc
\begin{tabular}{|llll|llll|} \hline
\multicolumn{8}{|c|}{$V_{{\rm eff}}(\rho,N)$} \\ \hline
\multicolumn{4}{|c}{\em {\rm Truncated Matrix}} & 
   \multicolumn{4}{c|}{\em {\rm Direct}} \\ \hline
$n$ & $\la$ & $N$ & $V_{{\rm eff}}(5,N)$ & $\ell$ & $|\ell_1|$ & 
$|\ell_2|$ & $V_{{\rm eff}}(5,N)$ \\ \hline
E & 0 & 0 & 0.011754666 & 0 & 0 & 0 & 0.011754562 \\
E & 0 & 2 & 0.037577818 & 1 & 0 & 0 & 0.037577462 \\
O & 0 & 2 & 0.000874927 & 0 & 1 & 1 & 0.000874911 \\
E & 0 & 4 & 0.062609805 & 2 & 0 & 0 & 0.062609219 \\
E & 0 & 4 & 0.00005971  & 0 & 2 & 2 & 0.00005971  \\
O & 2 & 4 & 0.00413519  & 1 & 1 & 1 & 0.00413512  \\
E & 1 & 1 & 0.024168    & 0 & 0 & 1 & 0.02416738  \\
E & 1 & 1 & 0.00029592  & 0 & 1 & 0 & 0.00029591  \\
O & 1 & 3 & 0.000024    & 0 & 2 & 1 & 0.00002426  \\
O & 1 & 3 & 0.00172537  & 0 & 1 & 2 & 0.00172529  \\
E & 1 & 3 & 0.050462    & 1 & 0 & 1 & 0.0504588   \\
E & 1 & 3 & 0.00226748  & 1 & 1 & 0 & 0.00226737  \\
E & 2 & 2 & 0.00000616  & 0 & 2 & 0 & 0.00000616  \\
O & 2 & 2 & 0.036849    & 0 & 0 & 2 & 0.03684737  \\
E & 2 & 4 & 0.000088636 & 1 & 2 & 0 & 0.000088629 \\
E & 2 & 4 & 0.062866    & 1 & 0 & 2 & 0.06286247  \\ \hline
\end{tabular}
\ec
	\caption{\it Some effective potential values in the $\overline{\delta}$ class}
\end{table}

\noindent
Here we must alert the reader. The above results were obtained without 
focusing on the permutational classifications. Some of these values  
belong to effective potentials which decay as inverse
logarithms for large $\rho$, others will decay much faster, as we will see.

\section*{Asymptotic Behaviour}

The matching of logarithmic derivatives provides a means of obtaining
information about the asymptotic behaviour of the eigenvalues, and
hence the effective potentials, as the hyper-radius, $\rho$, gets
large. There is however a particular difficulty in finding this behaviour.
It is that it is \underline{not} simply a case of looking at the
limiting behaviour of $_{2}F_{1}(a,b;c;\epsilon)$ and 
$_{2}F_{1}(a,b;c;1-\epsilon)$
as $\epsilon\rightarrow0$, because the expressions corresponding
to $a$ and $b$ both depend on $\rho$. 

We recall that the effective potential, $V_{{\rm eff}}$, is obtained by
matching, at the points $z=-1+\rho^{-2}$, the logarithmic derivatives
of the functions (R) and (L) given in the expressions (\ref{eq11},
\ref{eq21}), in conjunction with the Eqs. (\ref{nu1}, \ref{nu2}) given in the Appendix.

\subsection*{Inverse logarithmic behaviour}
In the analysis of the asymptotic behaviour of the adiabatic eigenvalues
in the Appendix, we show that, in the case corresponding to $\ell_{1}=0$,
the effective potential behaves as: 
\begin{equation}
\rho^{2}V_{\mbox{eff}}(\rho)=\frac{1}{A+B\ln\rho}\;+\frac{1}{4(N+1)^2 
\left(A+B\ln\rho\right)^{2}}\,.\label{eq:ab1}
\end{equation}
In this case $N$, the order of the harmonic, simplifies to $\left|\ell_2\right|+2\ell$
and the constants $A$ and $B$ are defined as 
\begin{equation}
A=\frac{1}{4(N+1)} \left[ \frac{2\, I_{0}\left(\sqrt{\frac{V_{0}}{2}}\right)
}{{\sqrt{\frac{V_{0}}{2}}}\, I_{1}\left(\sqrt{\frac{V_{0}}{2}}\right)}
-{\displaystyle \sum_{p=1}^{\ell}\frac{1}{p}}-{\displaystyle 
\sum_{p=1}^{\ell+\left|\ell_2\right|}\frac{1}{p}}+\ln2
\right]
\,,\qquad\qquad B= \frac{1}{2(N+1)}\;
.\label{eq:ab3}
\end{equation}
The $I_{i}$'s are modified Bessel functions of integer order of
the first kind, and it should be understood that 
\ensuremath{\sum_{1}^{0}\equiv0} . 

Equation (\ref{eq:ab1}) yields our very best description of the asymptotic 
behaviour of $V_{{\rm eff}}$, being accurate for the highest values of $\rho$, 
as well as for a wide range of lower values. We refer to the potential of
that description as $V_{{\rm best}}$.

Its leading term, together with the expressions
for the constants (\ref{eq:ab3}), was previously found by Klemm
and Larsen \cite{larsen-1}. It properly describes $V_{{\rm eff}}$ for the upper
range of values of $\rho$. We denote it as $V_{{\rm KL}}$. (Not to be confused with
the $KL$ of the hyperspherical basis,) 

A similar equation, giving an improved representation of the effective
potential for lower values of $\rho$, can be found by incorporating part 
of the quadratic term into the simple inverse logarithmic relationship.
We found as a result (see the Appendix):
\begin{equation}
\rho^{2}V_{\mbox{eff}}(\rho)\sim\frac{1}{A^*+B^* \ln\varrho},\label{eq:ab2}
\end{equation}
where
\begin{equation}
A^*= A-\frac{1}{4(N+1)^2}\,,\qquad{\cal \qquad}B^*=B.
\label{eq:ab4}
\end{equation}
We refer to it as $V_{{\rm wider}}$,  for the wider range of its description.

\esp
We emphasize that, in all 3 asymptotic expressions, the asymptotically 
dominant term in $1/(B \ln(\rho))$ always stays the same.

\esp
\begin{center}
	\begin{figure}[h]
\centering{}\centering{}
\includegraphics[scale=1.1] {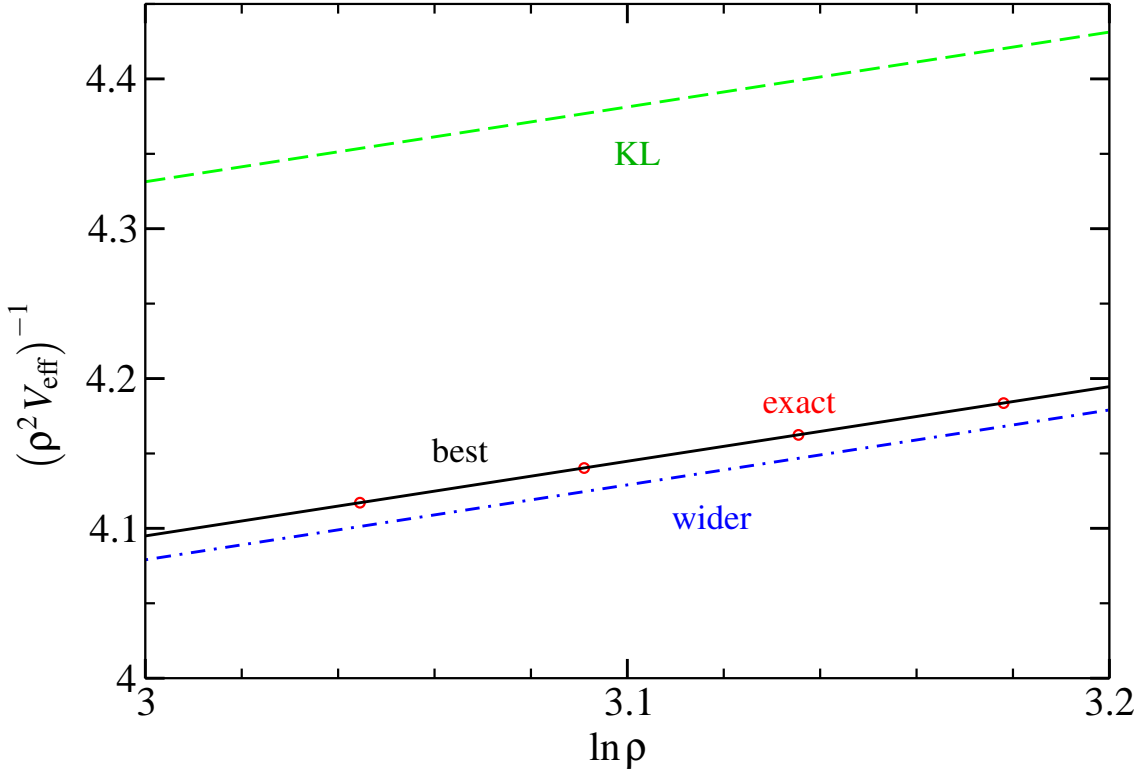}
\caption
{This figure shows the asymptotic behaviour of the effective potential
as a function of the hyperradius, for $\ell=\ell_{1}=\ell_2=0$, and 
compares the different approximations explained in this paper.	The 
red dots correspond to values of $V_{\rm exact}$, obtained by matching
numerically the functions (\ref{eq11}, \ref{eq21}); the black solid, 
to $V_{\rm best}$; the blue dot-dash line, to $V_{\rm wider}$ and the 
green dash line corresponds to $V_{\rm KL}$ \cite{larsen-1}
}
\end{figure}
\par\end{center}

In Figure (2), we present a comparison, in the $\rho$ large region, between the 
exact values of the effective potential (dots), $V_{{\rm exact}}$, with those 
obtained using the various asymptotic analytical expressions derived in the present paper. 
The case shown corresponds to the values of $\ell_{1}=0,\, \ell_2=0$ and 
$\ell=0$. The exact set of values was calculated numerically, by matching the 
logarithmic derivative of the solutions given in equation (\ref{twin}), at the point 
$z=-1+1/\rho^{-2}$. From the figure we can appreciate the inverse-logarithmic 
behaviour for the whole set of results, and that the exact results are 
very well described by the improved expression for the effective potential, 
$V_{{\rm best}}$.  

\esp
At this point, we would like to introduce another table ({\em Table 3}).

\esp
In it, we display values of the constant 
$\cal{A}$, in fits obtained by Zhen \cite{jei}
in her thesis work (and also published in \cite{zhen}). 
They are fits of the numerical values of some of the dominant effective 
potentials, for $\rho \geq 15$.
We join the comparable values of $A$ and $A^{*}$ of our analytic
expressions. \\
(In her thesis, she also compares
her $\cal B$'s with the postulated values of \cite{klemm}; sometimes her fits 
include a ${\cal C}/\rho^2$ in the inverse logarithmic expression.) 

\begin{table}[h]
\bc
\begin{tabular}{|lll|lll|c|c|c|} \hline
$n$ & $\la$ & $N$ & $\ell_1$ & $\ell$ & $\ell_2$ & ${\cal A}$ (Zhen) & 
{\bf $A$ } &  {\bf $A^*$ } \\ \hline
 $E$ & 0 & 0  &   0  &  0  &  0  &  2.6064        &    2.8293 & 2.5793  \\
 $E$ & 0 & 2  &   0  &  1  &  0  &  0.7581        &    0.7764 &  0.748659 \\
 $E$ & 0 & 4  &   0  &  2  &  0  &  0.4146        &    0.4159 &  0.4059 \\
 $E$ & 1 & 1  &   0  &  0  &  1  &  1.2381        &    1.2897 &  1.22715 \\
 $E$ & 1 & 3  &   0  &  1  &  1  &  0.5493        &    0.5511 &  0.5355 \\
 $E$ & 1 & 5  &   0  &  2  &  1  &  0.3356        &    0.3327 & 0.3257  \\  
\hline
\end{tabular}
\ec
\caption{\it Comparison of numerical and analytic \\
       asymptotic leading terms.}
\end{table}

We must emphasize that the values for ${\cal A}$, and ${\cal B}$, that were 
obtained by Zhen, correctly fit her data. 
We obtain similar results, when restricting ourselves to intermediate values 
of large $\rho$. \\ 
Our asymptotic expressions, first,  truly model the highest 
values of $\rho$, and then - with different degree of success - model 
the behaviour for smaller values of $\rho$.

\subsection*{Inverse $\rho$ behaviour}
We found, in the cases  $\ell_{1}\neq0$, that the potentials have a
different behaviour when $\rho$ is large. We show in the Appendix that
the leading term of the potential for this case, in the asymptotic
region, is

\begin{equation}
\rho^{2}V_{\mbox{eff}}(\rho) \simeq \frac{q}{\rho^{2\left|\ell_{1}
\right|}}\label{eq:ab5}
\end{equation}
where 
\begin{equation}
\frac{1}{q}=\frac{2^{|\ell_{1}|-2}}
{({N+1})C_{N-\ell}^{|\ell_{1}|}C_{|\ell_{1}|+\ell}^{|\ell_{1}|}}\left(\frac{1}{|\ell_{1}|}+\frac{{2}I_{|\ell_{1}|}\left(\sqrt{\frac{V_{0}}{2}}\right)}{\sqrt{{\frac{V_{0}}{2}}}\ I_{|\ell_{1}|+1}\left(\sqrt{\frac{V_{0}}{2}}\right)}\right), 
\ N= 2 \ell + \vert \ell_1 \vert +  \vert \ell_2 \vert .\label{eq:ab7}
\end{equation}
In figures (3a) and (3b) we exhibit this behaviour for the two cases, $\ell_{1}=1$
and $\ell_{1}=2$, respectively. In both examples $\ell_2=0$ and $\ell=0$.
The figures show how the exact values of the potentials are approaching their 
asymptotic behaviours, as given by Eq. (\ref{eq:ab5}).

\begin{figure}[h]
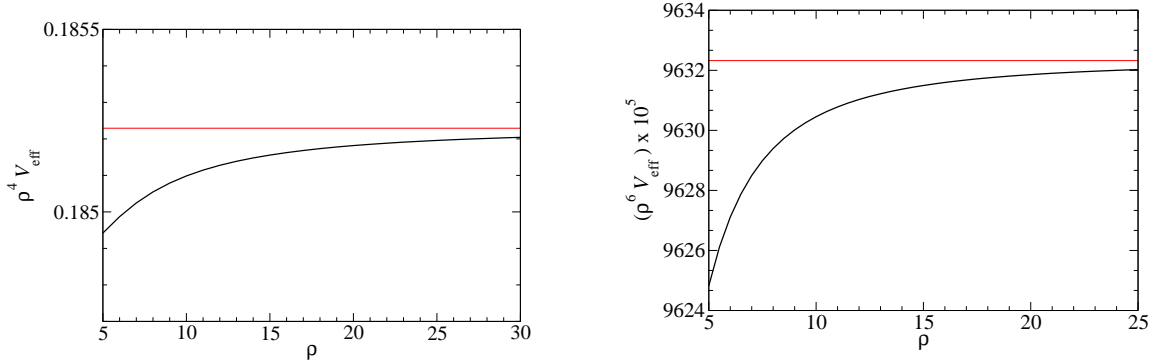

\centering{}
	\begin{subfigure}{.45\textwidth}
	\includegraphics[width=\textwidth] {fig5}
\caption
{This figure shows the effective potential $V_{{\rm exact}}$ (black 
curve) approaching its inverse-power asymptotic behaviour given by Eq. 
(\ref{eq:ab5}),	as $\rho$ increases, for $\ell_{1}=1$ and $\ell_2=0=\ell$. 
From Eq.  (\ref{eq:ab7}), $q=0.185229$ (straight red curve).
}
	\end{subfigure}
	\hspace{1cm}
	\begin{subfigure}{.45\textwidth}
		\includegraphics[width=\textwidth] {fig6}
\caption
{This figure shows the exact effective potential $V_{{\rm exact}}$ (black 
curve) , for  $\ell_{1}=2$ and $\ell_2=0=\ell$, approaching its inverse-power 
asymptotic behaviour given by Eq. (\ref{eq:ab5}), as $\rho$ increases. In this 
case, Eq. (\ref{eq:ab7}), gives a value for $q = 0.096323$ (straight red curve).
}
	\end{subfigure}
\caption
	{Asymptotic effective potentials for $\ell_{1}$ not equal to zero.}

\end{figure}




\newpage{}
\subsection*{Phase shifts}
In reference \cite{sigurd} \footnote{Unfortunately, an error crept in 
the ultimate equation.  We give the corrected result,
drawing from Eq.(\ref{yb}).}
it is shown that when $V_{{\rm eff}}$
behaves as $1/(\rho^2{\cal B} \ln\rho)$, the long-range
part of the potential dominates the behaviour of the phase shift,
and the latter goes to zero when $k\rightarrow0$ as
\begin{equation}
\delta \rightarrow\left(\frac{\pi}{4}\right)\,\frac{1}{{\cal B}\ln\, k}.
\label{eq:ab6}
\end{equation}
 
It is notable that this inverse logarithmic behaviour
occurs when $\ell_1$ equals zero, not $\ell_2$.
I.e., it is when the interacting pair, of particles, has zero angular
momentum. We are then drawn to the 2-body problem, in 2 dimensions, of 
interacting hard discs, where we find, for zero angular momentum, 
precisely the same result.

Indeed, for hard discs, and a potential of radius $\sigma$, 
the wave function $\psi$, in terms of Bessel and Neumann functions, 
and for angular momentum $L$, reads:
\bdis
\psi  \propto N_L(k\sigma) J_L(kr) - J_L(k\sigma) N_L(kr)
\edis
and thus
\bdis
\tan(\delta_L(k)) = \frac{J_L(k\sigma)} {N_L(k\sigma)}
\edis
At low energies, we therefore find that for hard discs, as 
for our step functions, that $\delta_0(k) \rightarrow (\pi/2) (1/\ln k\sigma)$,
and $\delta_L(k)  \propto  (k\sigma)^{2L}$, when $L \neq 0$.
. 

\esp
\esp
\noindent
We now wish to address the case of $\ell_1 \neq 0$. 

\esp
For $\ell_1 = 0$, we have argued that the `tail' of the effective potential 
dominates the small $k$ behaviour of the phase shift. We do so again when
$\ell_1 \neq 0$.

\esp
For a two-body problem with a radial potential that
falls off as $r^{-s}$ for large $r$, 
Mott and Massey \cite{Mott} obtain a behaviour
for the phase shift $\eta_{L}$ ,at low energies, of the form
\[{\displaystyle 
\lim_{k\rightarrow0}\, k^{s-2}\, \cot\,\delta_{L}= {\rm constant,} }
\]
when the angular momentum $L + \frac{1}{2} > (s-2)/2$. 
In our adiabatic 
hyperspherical  approach, the role of $L+1/2$, in the two-body system,
is played by $N+1$, and therefore the criterion for tail-dominant 
behaviour is that
$N+1$ be greater than $\left(s-2\right)/2$.
For $|\ell_1|$ in general (not zero), 
 we see that the minimum value of $N$ is $|\ell_1|$,
and that the value of $s$  equals $2 |\ell_1| + 2$. The inequality is always
satisfied $(|\ell_1| + 1) > |\ell_1|$ , and the tail always dominates.

As a result, we see that, given our asymptotic result of 
\begin{equation}
\rho^{2}V_{\mbox{eff}}(\rho) \simeq \frac{q}{\rho^{2\left|\ell_{1}
\right|}}
\end{equation}
and the expression of Mott and Massey,
our adiabatic phase shift result for $\ell_1$ is 
proportional to $(k\sigma)^{2 |\ell_1|}$,  and therefore in agreement 
with the hard disc result.

\subsection*{Symmetry considerations}

As representations of the permutational groups, 
the case $\overline{\delta}$ divides into
4 separate classes: symmetric and antisymmetric under the permutation of
the 2 interacting particles, and even and odd (gerade and ungerade) under
inversion of an additional operator.

In terms of the KL basis, the inverse logarithmic behaviour arises 
exclusively with the class of 2-body permutational symmetry, 
and $n$ even. 
We find also that upon expansion of the elements of the KL basis, in 
terms of the new basis, an element $Y^{\ell_1, \ell_2}_{\ell}$ with 
$\ell_1 = 0$ always appears. 
In contrast, such a term never appears in the other classes. In 
such cases  $\ell_1$ is always different from zero. 

(As an aside, we note that whenever $\ell_1$ is even, the element of the new 
basis will be symmetric under the 2-particle interchange.) 

\esp
From another point of view, the potential matrix elements, calculated with the
KL basis, 
always form  polynomials in $1/\rho^2$. However, in only one of the classes, 
the symmetric one with $n$ even, does the leading $1/\rho^2$ term appear. 
In all of the others, the leading term is of higher order, 
which implies a stronger decay of the effective potential for large $\rho$. 

\subsection*{Fully interacting system}
Here, perhaps, we can offer at least an intuitive insight into the inverse 
logarithmic behaviour of the fully interacting system. 

When all particles interact, $\ell_1$ and $\ell_2$ are no longer 
good quantum numbers, but the physical situation of close
approach of particles $1$ and $2$, associated with $\ell_1  = 0$, and 
now replicated in the other pairs,
must still take place, in the most symmetric wave function, and part of 
the wave function must reflect this 2-body situation.

\newpage
\section*{Conclusion}

We can now draw several important conclusions.

The first, and very important, is that, at least in the ($2 + 1$) case,  
the vast effort in establishing a hyperspherical basis, calculating matrix
elements, writing and perfecting complex numerical programs, seems to be
correct. \\
     It is now clear that the extensive numerical calculations of Zhen
\cite{jei}, and other authors \cite{klemm},  using the truncated matrix 
approach, provided good estimates of
the eigenvalues, the effective potentials, and the $2 + 1$
phase shifts of the third cluster.
 The results are consistent for the entire range of
values of $\rho$, taking into consideration the requirement for larger
$N_{{\rm max}}$ at larger values of $\rho$.

We were also able to demonstrate the all important logarithmic behaviour 
in the asymptotic form of some of the effective potentials, which so 
caught our eye, and which we tried to characterize in \cite{klemm}. 
This insures
that the corresponding phase shifts (dominant at low energies)
go to zero, as the wave number goes to zero.
For the other $2 + 1$ phase shifts, characterized by other 
group classifications of the harmonics, we can demonstrate by explicit
calculations that both the asymptotic form of the effective potentials and
the phase shifts go to zero in a stronger manner.

In fact, we were able to offer very complete and beautiful 
asymptotic expressions for all the cases involved in $\overline{\delta}$, and
present accurate numerical calculations for all the effective potentials, 
and for all desired values of $\rho$. 

\esp
We would love to obtain similar asymptotic expressions for the effective
potentials of the fully interacting problem. If we were able to do this, it
would simplify enormously the cluster calculations, as well as increase
their accuracy.

\vspace{0.5cm}

\section*{Acknowledgements} All the authors acknowledge, with thanks,
support from their respective Universities and Institutes, not only for 
themselves, but for extending support and hospitality to the other authors 
of this paper. At Deakin University we thank a `Bilateral Science and 
Technology Program (Australia,  US) for a grant, and a study leave 
from the University.

S.Y. Larsen would like to pay tribute to John E. Kilpatrick, whose help was
essential in calculating the potential matrix elements, but whose
work was, unfortunately, never published. It is referred to as 
reference (3), in \cite{zhen}.

\vspace{.5cm}

\newpage{}

\section{Appendix}

This Appendix is devoted to the asymptotical ($\rho$ infinite)
behaviour of the effective potential $V_{{\rm eff}}(\rho)$.  
We recall that the effective potential $V_{{\rm eff}}(\rho)$ is obtained by
matching, at the points $z=-1+\rho^{-2}$, the logarithmic derivatives
of the functions (R) and (L), given in the expressions {[}\ref{eq11},
\ref{eq21}{]} . Also 
that the derivative of a hypergeometrical function reads : 
\begin{equation}
\frac{d}{dy}\,\left[ {}_{2}F_{1}(a,b;c;y) \right] = \frac{ab}{c}\,{}_{2}F_{1}(a+1,b+1;c+1;y)  \,
\quad  c \ne 0 \,.\label{eq:derf}
\end{equation}
We note that, for our effective potentials, our $\nu_1$ and $\nu_2$ take the following form:

\begin{eqnarray}
\nu_{1} & = & -\frac{1+|\ell_{1}|+|\ell_{2}|}{2}+\sqrt{\frac{(1+2\ell+
|\ell_{1}|+|\ell_{2}|)^{2}}{4}+\frac{\rho^{2}V_{{\rm eff}}
(\rho)}{4}}\label{nu1}\\
 & \hspace{1cm} & \mbox{and}\nonumber \\
\nu_{2} & = & -\frac{1+|\ell_{1}|+|\ell_{2}|}{2}+\sqrt{\frac{(1+2\ell+|\ell_{1}|+|\ell_{2}|)^{2}}{4}+\frac{\rho^{2}V_{{\rm eff}}(\rho)}{4}-\frac{\rho^{2}V_{0}}{4}}\,.\label{nu2}
\end{eqnarray}

\subsection{Asymptotic expression for the Logarithmic derivative of the 
lefthand part (L)}

At the matching point $z=-1+1/\rho^{2}$, the hypergeometric series
$_{2}F_{1}(-\nu_{2},\nu_{2}+|\ell_{1}|+|\ell_{2}|+1;|\ell_{1}|+1;\frac{1}{2}(1+z))$
has two arguments expected to be infinite as $\rho\to\infty$, since
$\nu_{2}\simeq \imath \,\rho\sqrt{V_{0}}/2$. Also, the argument $z$ is close
to $-1$, at $\rho$ large. Following the argument developed in the
Ref. \cite{erd1a} we write : 
\begin{equation}
_{2}F_{1}(-\nu_{2},\nu_{2}+|\ell_{1}|+|\ell_{2}|+1;|\ell_{1}|+1;\frac{1}{2}(1+z))\simeq\sum_{n=0}^{\infty}\left(\frac{-\nu_{2}^{2}(1+z)}{2}\right)^{n}\frac{1}{(|\ell_{1}|+1)_{n}n!}\vert_{z=-1+1/\rho^{2}}\,.\label{eq:}
\end{equation}
When $\rho\to\infty$ the latter hypergeometric becomes : 
\begin{eqnarray*}
 \lim_{\rho \to \infty}\left[ _{2}F_{1}\left(-\nu_{2},\nu_{2}+|\ell_{1}|+|\ell_{2}|+1;|\ell_{1}|+1;
\frac{1}{2\rho^{2}}\right)\right] 
&  = & \sum_{n=0}^{\infty}\left(\frac{V_{0}}{8}\right)^{n}
\frac{1}{(|\ell_{1}|+1)_{n}n!} \nonumber\\
 & = & |\ell_{1}|!\left(\frac{8}{V_{0}}\right)^
{|\ell_{1}|/2}I_{|\ell_{1}|}\left(\sqrt{\frac{V_{0}}{2}}\right)\ . 
\end{eqnarray*}
We then again follow \cite{erd1a} and, using (\ref{eq:derf}),  
when $\rho \rightarrow \infty$,
we write : 
\begin{eqnarray}
 &  & \rho^{-2}\frac{d}{dz} \left[{}_{2}F_{1}(-\nu_{2},\nu_{2}+
|\ell_{1}|+|\ell_{2}|+1;|\ell_{1}|+1;\frac{1}{2}(1+z)) \right] \nonumber \\
 & \simeq & \frac{-\nu_{2}(\nu_{2}+|\ell_{1}|+|\ell_{2}|+1)}{2(|\ell_{1}|+1)}\sum_{n=0}^{\infty}\left(\frac{-\nu_{2}^{2}(1+z)}{2}\right)^{n}\frac{1}{(|\ell_{1}|+2)_{n}n!}\vert_{z=-1+1/\rho^{2}}\nonumber \\
 &\underset{\rho \to \infty}{\longrightarrow}  & \sqrt{\frac{V_{0}}{8}\ }|\ell_{1}|!\left(\frac{8}{V_{0}}\right)^{|\ell_{1}|/2}I_{|\ell_{1}|+1}\left(\sqrt{\frac{V_{0}}{2}}\right)\,.
\end{eqnarray}

We conclude that the logarithmic derivative of the function $_{2}F_{1}$,
with respect to $z$, satisfies, at the matching point $z=-1+1/\rho^{2}$,
\begin{equation}
	\lim_{\rho\to\infty}\rho^{-2}\frac{d}{dz}\ln({}_{2}F_{1}(-\nu_{2},\nu_{2}+|\ell_{1}|+|\ell_{2}|+1;|\ell_{1}|+1;1/(2\rho^{2})))=\sqrt{\frac{V_{0}}{8}}I_{|\ell_{1}|+1}\left(\sqrt{\frac{V_{0}}{2}}\right)/I_{|\ell_{1}|}\left(\sqrt{\frac{V_{0}}{2}}\right)\,.\label{eq:hyperlim}
\end{equation}

\subsection{Asymptotic expressions of the righthand part (R)}

We now consider the equations (\ref{eq11}, \ref{nu1}), concerning
the part (R), and identify the hypergeometric series $_{2}F_{1}(-\nu_{1},\nu_{1}+|\ell_{1}|+|\ell_{2}|+1;|\ell_{2}|+1;\frac{1}{2}(1-z))$
with $_{2}F_{1}(a,b,a+b-m,z^{*})$ of \cite{erd1b}. We then have : 
\begin{eqnarray}
	&  & _{2}F_{1}(-\nu_{1},\nu_{1}+|\ell_{1}|+|\ell_{2}|+1;|\ell_{2}|+1;\frac{1}{2}(1-z))={}_{2}F_{1}(a,b,a+b-m,z^{*})\nonumber \\
 &  & a=-\nu_{1}\ ,\qquad\quad b=\nu_{1}+|\ell_{1}|+|\ell_{2}|+1\nonumber \\
 &  & a+b-m=|\ell_{2}|+1=-\nu_{1}+\nu_{1}+|\ell_{1}|+|\ell_{2}|+1-m=|\ell_{1}|+|\ell_{2}|+1-m\ ,\nonumber \\
 &  & z^{*}=\frac{1-z}{2}\ .\label{eqref}
\end{eqnarray}
The above equations imply that $m=|\ell_{1}|$. 
For values of $z^{*}$ close to unity, the leading terms of Eq.(\ref{eqref}) 
are given by (see \cite{erd1b})
\begin{eqnarray}
_{2}F_{1}(a,b,a+b-m,z^{*}) & \simeq & \frac{\Gamma[a+b-m]\ 
\Gamma[m]}{\Gamma[a] \, \Gamma[b]}(1-z^{*})^{-m} \nonumber  \\
 & +  & \frac{(-1)^{m} \ \Gamma[a+b-m]\ }{\Gamma[a-m]\ \Gamma[b-m] m!}(h_{0}-\ln(1-z^{*})) \nonumber \\
 &  & \qquad \qquad \qquad \qquad \qquad a,b\ne0,-1,-2,-3,...\label{eqrefE0} 
\end{eqnarray}
in terms of the function $\psi(z)=\frac{d\ln(\Gamma(z)}{dz}$, and as
$\rho$ goes to $\infty$.
In Eq. (\ref{eqrefE0}) $h_{0}$ denotes 
\begin{equation}
h_{0}=\psi(1) + \psi(1+m) -\psi(a)-\psi(b) \ . \label{eq:hn}
\end{equation}
Guided by Eqs. (\ref{eqrefE0}) we have 
\begin{eqnarray}
_{2}F_{1}(-\nu_{1},\nu_{1}+|\ell_{1}|+|\ell_{2}|+1;|\ell_{2}|+1;\frac{1}{2}(1-z))
 & \simeq &  \frac{\Gamma[|\ell_{2}|+1]\Gamma[|\ell_{1}|]}{\Gamma[-\nu_{1}]
\Gamma[\nu_{1}+|\ell_{1}|+|\ell_{2}|+1]}\left(\frac{1}{2}+\frac{z}{2}
\right)^{-\left|\ell_{1}\right|}\nonumber \\
 & + & \frac{(-1)^{|\ell_1|} \ \Gamma[|\ell_{2}|+1]}{\Gamma[-\nu_{1} - |\ell_1|]
\Gamma[\nu_{1}+|\ell_{2}|+1]|\ell_1|!}\left(h_{0}-\ln\left(\frac{1+z}{2}\right)\right)
\nonumber \\
 &  &  \qquad \qquad\qquad  \quad\nu_{1}\ne0,1,2,...\label{eq1p}
\end{eqnarray}
and
\begin{equation}
h_{0}=\psi(1) + \psi(1+|\ell_1|)- \psi(-\nu_{1})-\psi(\nu_{1}+ |\ell_1| +|\ell_{2}|+1)\ .
\end{equation}
The r.h.s. of each expression in Eq. (\ref{eq1p}) involve $\Gamma[-\nu_{1}]$
or $\Gamma[- \nu_1 + |\ell_1|]$,
which are singular for every value of the argument equal to an integer. 
For $\nu_1$ close to $\ell$, we can write \cite{erd1c} 
\begin{eqnarray}
\Gamma[-\nu_{1} - |\ell_1|] & = & -\frac{\pi}{\Gamma[\ell + |\ell_1|]
(\ell + |\ell_1|) \sin(\pi(\nu_{1} + |\ell_1|))}
\vert_{\nu_{1}=\ell+\nu_{1}-\ell} \nonumber \\
 &=& (-)^{\ell + |\ell_1|+1}
\frac{\pi}{\Gamma[\ell+|\ell_1|+1]\sin\pi(\nu_{1}-\ell)}\nonumber \\
 & \simeq & (-1)^{\ell+|\ell_1| +1}\frac{1}{\Gamma[\ell+|\ell_1|+1]}
\frac{1}{\nu_{1}-\ell}\ +{\cal O}(\nu_{1}-\ell)\,.\label{nu1l}
\end{eqnarray}
which is also valid when $|\ell_1|$ equals zero.

\subsection{Case $\ell_{1}=0$. }

We first analyze this case because the corresponding asymptotic behaviour
differs from that when $\ell_1$ is different from zero, and displays the 
inverse logarithmic behaviour which first attracted our attention.

\subsubsection{Logarithmic derivative of the part (R)}

Taking into account equation (\ref{eq:derf}) and the second term of 
(\ref{eq1p}) we  evaluate the inverse of the logarithmic derivative of 
part (R), Eq. (\ref{eq11}).
\begin{eqnarray}
	&  & \rho^{2}\left[\frac{d}{dz}\ln{}(_{2}F_{1}(-\nu_{1},\nu_{1}+|\ell_{1}|+|\ell_{2}|+1;|\ell_{2}|+1;\frac{1}{2}(1-z)))\vert_{z=-1+1/\rho^{2}}\right]^{-1}\nonumber \\
 & \simeq & -h_{0}-\ln(2)-2\ln(\rho)\,.\label{eq2p0}
\end{eqnarray}

When $\rho$ is large, $\rho^{2}V_{\rm eff}\left(\rho\right)$ is expected
to be small \cite{klemm} and $\nu_{1}$ to be in the vicinity of $\nu_{1}=\ell$,  $\ell$ integer.
We then have : 
\begin{eqnarray}
h_{0}=-2\gamma-\psi(-\nu_{1})-\psi(\nu_{1}+\vert\ell_{2}\vert+1) & \simeq & -\frac{1}{\nu_{1}-\ell}-H_{\ell}-H_{\ell+\vert\ell_{2}\vert}+{\cal O}(\nu_{1}-\ell)\,,\label{h0}
\end{eqnarray}
where $\gamma$ is the Euler's constant. In Eq. (\ref{h0}) $H_{n}$
denotes : 
\begin{equation}
H_{n}=\sum_{p=1}^{n}\frac{1}{p},\qquad n\geq1,\qquad H_{0}=0\ .\label{eq:Hn}
\end{equation}

\subsubsection{Matching parts (R) and (L) for $\rho$ large.}

The logarithmic derivative of part (L) is given by Eq. (\ref{eq:hyperlim}).
For $\ell_{1}=0$, we then have: 
\begin{equation}
	\lim_{\rho\to\infty}\rho^{2}\left[\frac{d}{dz}\ln{}(_{2}F_{1}(-\nu_{2},\nu_{2}+|\ell_{1}|+|\ell_{2}|+1;|\ell_{1}|+1;1/(2\rho^{2})))\right]^{-1}=I_{0}\left(\sqrt{\frac{V_{0}}{2}}\right)\sqrt{\frac{8}{V_{0}}}/I_{1}\left(\sqrt{\frac{V_{0}}{2}}\right)\,.\label{eq:hyperlim0}
\end{equation}
Taking into account the results of section (1.2), we finally have,
from the matching of the left and righthand parts, the following expression:
\begin{equation}
\frac{\sqrt{8}\ I_{0}\left(\sqrt{\frac{V_{0}}{2}}\right)}{\sqrt{V_{0}}\ I_{1}\left(\sqrt{\frac{V_{0}}{2}}\right)}\simeq\frac{1}{\nu_{1}-\ell}+H_{\ell}+H_{\ell+|\ell_{2}|}-\ln(2)-2\ln(\rho)\ .\label{eq:final0}
\end{equation}
We solve the above, to obtain an asymptotic expression for $\nu_1$, i.e., for 
$\nu_{1}\simeq\nu_{1}^{a}$, where
\begin{equation}
\nu_{1}^{a}=\ell+\frac{1}{\frac{\sqrt{8}\ I_{0}\left(\sqrt{\frac{V_{0}}{2}}\right)}{\sqrt{V_{0}}\ I_{1}\left(\sqrt{\frac{V_{0}}{2}}\right)}-H_{\ell}-H_{\ell+|\ell_{2}|}+\ln(2)+2\ln(\rho))},\label{eq:nuanap}
\end{equation}
and consequently 
\begin{eqnarray}
\nu_{1}^{a}-\ell & = & \frac{1}{\tilde{A}+\tilde{B}\ln(\rho)}\label{nuanap1}\\
\tilde{A} & = & \frac{\sqrt{8}\ I_{0}\left(\sqrt{\frac{V_{0}}{2}}\right)}{\sqrt{V_{0}}\ I_{1}\left(\sqrt{\frac{V_{0}}{2}}\right)}-
H_{\ell}-H_{\ell+|\ell_{2}|}+\ln(2)\ ,
\quad \tilde{B}=2\ .\label{nuanap2}
\end{eqnarray}

\subsubsection{Effective potential.}

The solution of Eq. (\ref{nu1}) for the potential $V_{{\rm eff}}(\rho)$
is given by : 
\begin{equation}
\frac{\rho^{2}V_{{\rm eff}}(\rho)}{4}=(\nu_{1}-\ell)^{2}+(1+N)(\nu_{1}-\ell) ,
\ \  N=2\ell+|\ell_{2}|\,.\label{eq:Vana}
\end{equation}
If we take into account the fact that $\nu_{1}\simeq\nu_{1}^{a}$
and that $\rho^{2}V_{{\rm eff}}(\rho)$ is small, and neglect the quadratic term,
we can then write
\begin{equation}
\nu_{1}^{a}-\ell\simeq\frac{\rho^{2}V_{{\rm eff}}(\rho)}{4(1+N)}+{\cal O}(\rho^{2}V_{{\rm eff}}(\rho))\,,\label{eq:nu1ml}
\end{equation}
Using Eqs. (\ref{nuanap1}, \ref{nuanap2}) we obtain
the result found by Klemm and Larsen \cite{larsen-1}: 
\begin{equation}
\rho^{2}V_{{\rm eff}}(\rho)\simeq\rho^{2}V_{{\rm KL}}(\rho)=
\frac{1}{A+B\ln(\rho)}\,,\label{eq:param0}
\end{equation}
where : 
\begin{equation}
A = \frac{\tilde{A}}{4(N+1)} , \qquad \qquad 
B = \frac{\tilde{B}}{4(N+1)} \label{AB}
\end{equation}

To arrive at this expression we neglected a term in $1/(\ln(\rho)^{2}$.
We obtain a more accurate expression for the effective potential,
for a wider range of $\rho$ large, by including the term. Using
 (\ref{eq:Vana}), we obtain:

\begin{eqnarray}
\rho^{2}V_{\rm best}(\rho) & = 
& \frac{1}{4(N+1)^2(A+B\ln(\rho))^{2}}+
\frac{1}{A+B\ln(\rho)}\,.\label{eq:Vacc}
\end{eqnarray}
This asymptotic expression provides the best representation 
of our accurate numerical data.

We now show
that we can partially include the quadratic term in an 
equation similar to Eq. (\ref{eq:param0}),
with an inverse linear logarithmic behaviour for the potential, but
with a parameter $A^{*}$ smaller than the parameter $A$ found above, and
in ref. \cite{larsen-1}, and giving a better fit for lower values of $\rho$.
We proceed as follows: 
\begin{eqnarray}
\rho^{2}V_{\mbox{eff}}(\rho) & =&\frac{1}{A+B\ln\rho}\,
\left[1+\frac{1}{4\left(N+1\right)^2\left(A+B\ln\rho\right)}\right] 
\nonumber\\
& =  &\frac{1}{A+B\ln\rho}\,\left[\frac{1}
{1-\left(1/\left[1+4\left(N+1\right)^2\left(A+B\ln\rho\right)\right]
\right)}\right] 
\nonumber\\
 & \simeq  &\frac{1}{\left(A+B \ln\rho - \frac{1}{4 (N+1)^2}\right)}. 
\nonumber
\end{eqnarray}
Thus, the effective potential can be approximated by the equation,
\begin{equation}
\rho^{2}V_{\mbox{eff}}(\rho)\sim\frac{1}{A^*+B^* \ln\varrho},\label{eq:ab8}
\end{equation}
where
\begin{equation}
A^*= A-\frac{1}{4(N+1)^2}\,,\qquad{\cal \qquad}B^*=B.
\label{eq:ab9}
\end{equation}
We note that the ultimate term, in 
$1/(B\ln(\rho))$, always stays the same. 

\subsection{Case $\ell_{1}\protect\neq0$}

\subsubsection{Inverse powers of $\rho$}

For this case, we will need to consider both terms occurring in 
Eq. (\ref{eq1p}).  \\
The first term is divergent  approaching $z=-1$ ( or $\rho$ infinite), 
for fixed $\nu_1$,  since it behaves like
$\rho^{|2\ell_{1}|}$ as $\rho$ $\to$ $\infty$. We compensate for this by 
adjusting $\nu_1$.  
Since we expect the potential $\rho^2 \, V_{{\rm eff}}(\rho)$ to
tend to zero for $\rho$ infinite \cite{klemm}, $\nu_{1}$ will
be close to an integer $\ell$. 
To keep the term $ \rho^{2|\ell_{1}|}/\Gamma(-\nu_1)$ (and therefore
   $\rho^{2|\ell_{1}|}(\nu_{1}-\ell)$) finite, we set
\begin{equation}
\nu_{1}\simeq\ell+\frac{q}{4(1+N)}\ \rho^{-2|\ell_{1}|}\ +
{\cal O}(\rho^{-2|\ell_{1}| - 2})\ ,\label{eq:nu1t}
\end{equation}
where the parameter $q$ is expected to be a constant to be determined.
Taking into account Eq. (\ref{eq:nu1ml}), we find 
\begin{equation}
\nu_{1}-\ell\simeq\frac{\rho^{2}V_{{\rm eff}}(\rho)}{4(1+N)},
\quad N=2\ell+|\ell_{1}|+|\ell_{2}|\,.\label{eq:eqV}
\end{equation}
and therefore:
\begin{equation}
\rho^{2}V_{{\rm eff}}(\rho)\simeq\frac{q}{\rho^{2|\ell_{1}|}}\ .\label{eq:potVa}
\end{equation}

\subsubsection{Case $\ell=0$. Derivative of part (R)}

Let us first consider the case where $\ell=0$ for the sake of simplicity. \\
We will need to evaluate both terms of Eq. (\ref{eq1p}). 

Knowing the result (!), we start with the second term.

The important contribution in the numerator comes from $- \psi(-\nu_1)$ 
in $h_0$. In the denominator, it is $\Gamma(-\nu_1 - |\ell_1|)$ that prevails. 
Both functions  give contributions that behave as $\rho^{2 |\ell_1|}/ q$. 
Taking the 
ratio, and account of the various factors, we obtain a total result of $1$ 
for the second term. 

Proceeding with the first term, we evaluate it, as we have outlined above. 
We then have:

\begin{eqnarray}
_{2}F_{1}(-\nu_{1},\nu_{1}+|\ell_{1}|+|\ell_{2}|+1;|\ell_{2}|+1;1-1/(2\rho^{2})) & \simeq & 1-\nu_{1}\frac{\Gamma[|\ell_{2}|+1]\Gamma[|\ell_{1}|]}{\Gamma[|\ell_{1}|+|\ell_{2}|+1]}\left(2\rho^{2}\right)^{|\ell_{1}|}\nonumber \\
 & \underset{\rho \to \infty}{\longrightarrow} & 1-\frac{a_{|\ell_{1}|,|\ell_{2}|,0}}{|\ell_{1}|}q\ ,\label{hyper20}
\end{eqnarray}
where 
\begin{equation}
a_{|\ell_{1}|,|\ell_{2}|,0}=2^{|\ell_{1}|-2}((1+N)C_{N}^{|\ell_{1}|})^{-1}
\ ,\label{eq:a0}
\end{equation}
where $C_{n}^{p}$ denotes
the binomial coefficient $C_{n}^{p}=n!/(p!(n-p)!)$. Higher order terms start 
with order $\ln(\rho)/\rho^2$. 

We now evaluate the derivative, with respect to z, of our hypergeometrical 
function. This is given by equation (\ref{eq:derf}), together with 
equation (\ref{eqrefE0}), but taking care that $m=|\ell_{1}|+1$. 
This time our second term will not contribute, and we will have:
\begin{eqnarray}
 &  & \frac{d}{dz}\left[{}_{2}F_{1}(-\nu_{1},\nu_{1}+|\ell_{1}|+
|\ell_{2}|+1;|\ell_{2}|+1;\frac{1}{2}(1-z))\right]_{z=-1+1/\rho^{2}}
\nonumber \\
 & \simeq & \nu_{1}\frac{(1+|\ell_{1}|+|\ell_{2}|)}{2(1+|\ell_{2}|)}
\frac{\Gamma[|\ell_{2}|+2]\Gamma[|\ell_{1}|+1]}{\Gamma[|\ell_{1}|+
|\ell_{2}|+2]}\left(\frac{1}{2\rho^{2}}\right)^{-|\ell_{1}|-1}\nonumber \\
 & \simeq & \rho^{2}q\ a_{|\ell_{1}|,|\ell_{2}|,0}\ .\label{deriv01}
\end{eqnarray}

\subsubsection{Case $\ell\protect\neq0$. Derivative of part (R)}

For $\ell \neq 0$, we now proceed in exactly the same fashion, as we did 
for $\ell = 0$. The second term in our evaluation of ${}_2F_1$ again gives 
us a constant, and the first term gives us a term in $q$.
For $\rho$ large and $\nu_1$ close to $\ell$, we then find:
\begin{eqnarray}
_{2}F_{1}(-\nu_{1},\nu_{1}+|\ell_{1}|+|\ell_{2}|+1;|\ell_{2}|+1;1-1/(2\rho^{2})) & \simeq & (-)^{\ell}\frac{(|\ell_{1}|+\ell)!|\ell_{2}|!}{(|\ell_{2}|+\ell)!|\ell_{1}|!}\nonumber \\
 & + & (-)^{\ell+1}(\nu_{1}-\ell)\frac{\Gamma[\ell+1]\Gamma[|\ell_{2}|+1]\Gamma[|\ell_{1}|]}{\Gamma[|\ell_{1}|+|\ell_{2}|+\ell+1]}\left(2\rho^{2}\right)^{|\ell_{1}|}\nonumber \\
 & \underset{\rho \to \infty}{\longrightarrow} & (-)^{\ell}\frac{(|\ell_{1}|+\ell)!|\ell_{2}|!}{(|\ell_{2}|+\ell)!|\ell_{1}|!}\left(1-\frac{a_{|\ell_{1}|,|\ell_{2}|,\ell}}{|\ell_{1}|}q\right))\ .\label{hyper2}
\end{eqnarray}
The coefficient $a_{|\ell_{1}|,|\ell_{2}|,\ell}$ is given
by : 
\[
a_{|\ell_{1}|,|\ell_{2}|,\ell}=2^{|\ell_{1}|-2}((1+N)\ C_{N-\ell}^{|\ell_{1}|}\ C_{|\ell_{1}|+\ell}^{|\ell_{1}|})^{-1}\ .
\]

Again, and for the same reasons, we obtain:
\begin{eqnarray}
 &  & \frac{d}{dz}\left[{}_{2}F_{1}(-\nu_{1},\nu_{1}+|\ell_{1}|+
|\ell_{2}|+1;|\ell_{2}|+1;\frac{1}{2}(1-z))\right]_{z=-1+1/(2\rho^{2})}=\frac{\nu_{1}(1+|\ell_{1}|+|\ell_{2}|+\ell)}{2(1+|\ell_{2}|)}\nonumber \\
 &  & \times{}_{2}F_{1}(1-\nu_{1},\nu_{1}+|\ell_{1}|+|\ell_{2}|+2;|\ell_{2}|+2;1-1/(2\rho^{2}))\nonumber \\
 &  & \simeq(-)^{\ell}(\nu_{1}-\ell)\Gamma[\ell + 1]\frac{(1+|\ell_{1}|+|\ell_{2}|+\ell)}{2(1+|\ell_{2}|)}\frac{\Gamma[|\ell_{2}|+2]\Gamma[|\ell_{1}|+1]}{\Gamma[\ell+|\ell_{1}|+|\ell_{2}|+2]}\left(\frac{1}{2\rho^{2}}\right)^{-|\ell_{1}|-1}\nonumber \\
 & & \simeq \rho^{2} q \ (-)^{\ell}\frac{(|\ell_{1}|+\ell)!|\ell_{2}|!}
{(|\ell_{2}|+\ell)!|\ell_{1}|!}\ a_{|\ell_{1}|,|\ell_{2}|,\ell}
\label{hyper2p}
\end{eqnarray}

\subsubsection{Matching (R) and (L),  and our asymptotic result 
for the effective potential.}

We obtain, for the logarithmic derivative of the righthand side:
\begin{equation}
\lim_{\rho \to \infty} \rho^{-2}\frac{d}{dz}\left[ \ln[{}_{2}F_{1}(-\nu_{1},\nu_{1}+|\ell_{1}|+
|\ell_{2}|+1;|\ell_{2}|+1;\frac{1}{2}(1-z))]\right]_{z=-1+1/\rho^{2}}
= \frac{a_{|\ell_{1}|,|\ell_{2}|,\ell}\ q}{1-\frac{a_{|\ell_{1}|,
|\ell_{2}|,\ell}}{|\ell_{1}|}\ q}\,.
\label{eq:eq32}
\end{equation}
Matching this result with the corresponding
result (\ref{eq:hyperlim}) for the lefthand side, we obtain the analytical
value of $q$. 
\begin{equation}
\frac{1}{q}=\frac{2^{|\ell_{1}|-2}}{(1+N)C_{N-\ell}^{|\ell_{1}|}C_{|\ell_{1}|+\ell}^{|\ell_{1}|}}\left(\frac{1}{|\ell_{1}|}+\frac{2\sqrt{2}I_{|\ell_{1}|}\left(\sqrt{\frac{V_{0}}{2}}\right)}{\sqrt{V_{0}}\ I_{|\ell_{1}|+1}\left(\sqrt{\frac{V_{0}}{2}}\right)}\right)
\end{equation}
and therefore the analytic expression for the asymptotic behaviour
of the effective potential: 
\begin{equation}
V_{{\rm eff}}(\rho)=\frac{q}{\rho^{2|\ell_{1}|+2}}\,.\label{eq:potana}
\end{equation}

\end{document}